\documentclass[useAMS]{mn2e}

\usepackage{times}
\usepackage{epsfig}

\newcommand{\etal}{\rm et~al.}

\def\lesssim{\mathrel{\hbox{\rlap{\hbox{\lower4pt\hbox{$\sim$}}}\hbox{$<$}}}}
\def\gtrsim{\mathrel{\hbox{\rlap{\hbox{\lower4pt\hbox{$\sim$}}}\hbox{$>$}}}}

\title[Radio Emissions from Companions to Giants]
{Radio Emissions from
Substellar Companions of Evolved Cool Stars}

\author[Ignace, Giroux, Luttermoser]
{
Richard Ignace, Mark L.\ Giroux, \& Donald G.\ Luttermoser \\
       Department of Physics \& Astronomy,
       East Tennessee State University,
       Box 70652,
       Johnson City, TN 37614
       USA
}

\begin{document}

\maketitle

\begin{abstract}

A number of substellar companions to evolved cool stars have now
been reported.  Cool giants are distinct from their progenitor Main
Sequence (MS) low-mass stars in a number of ways.  First, the mass
loss rates of cool giant stars are orders of magnitude greater than
for the late-type MS stars.  Second, on the cool side of the
Linsky-Haisch ``dividing line'', K and M giant stars are not X-ray
sources, although they do show evidence for chromospheres.  As a
result, cool star winds are largely neutral for those spectral types,
suggesting that planetary or brown dwarf magnetospheres will not
be effective in standing off the stellar wind.  In this case one
expects the formation of a bow shock morphology at the companion,
deep inside its magnetosphere.  We explore radio emissions from
substellar companions to giant stars using (a) the radiometric
Bode's law and (b) a model for a bow shock morphology.  Stars that
are X-ray emitters likely have fully ionised winds, and the radio
emission can be at the milli-Jansky level in favorable conditions.
Non-coronal giant stars produce only micro-Jansky level emissions
when adjusted for low-level ionisations.  If the largely neutral
flow penetrates the magnetosphere, a bow shock results that can be
strong enough to ionise hydrogen.  The incoherent cyclotron emission
is sub-microJansky.  However the long wavelength radio emission of
solar system objects is dominated by the cyclotron maser instability
(CMI) mechanism.  Our study leads to the following two observational
prospects.  First, for coronal giant stars that have ionised winds,
application of the radiometic Bode's law indicates that long wavelength
emission from substellar companions to giant stars may be detectable or
nearly detectable with existing facilities.  Second, for the non-coronal
giant stars that have neutral winds, the resultant bow shock may act
as a ``feeder'' of electrons that is well-embedded in the companion's
magnetosphere.  Incoherent cyclotron emissions are far too faint to be
detectable, even with next generation facilities; however, much brighter
flux densities may be achievable when CMI is considered.

\end{abstract}


\section{INTRODUCTION}

The discovery of extrasolar planets around solar-type stars has now
become a fairly regular occurrence (e.g., see the Extrasolar Planets
Encyclopaedia maintained by J.~Schneider at
www.obspm.fr/encycl/encycl.html).  In addition to these, there have
now been numerous detections of planetary companions to cool {\em
giant} stars, (early detections include Frink \etal\ 2002, Setiawan
\etal\ 2003, and Sato \etal\ 2003).  With the ansatz that the
formation of planetary systems around solar type stars is a relatively
common occurrence, plus the recognition that such stars will evolve
to become giants that experience significant mass loss on the way
to becoming white dwarf stars, it becomes interesting to consider
how giant star winds might affect substellar companions.

This paper is not the first to entertain questions about the eventual
evolution of planetary systems during late stellar phases.  Several
authors have considered the angular momentum transfer between the
orbit of a planet or Brown Dwarf with a red giant and/or its wind
(Livio \& Soker 1984; Livio, Soker, \& Harpaz 1984; Livio \& Soker
2002).  Such effects could lead to ``sculpting'' of Planetary Nebulae
(e.g., Soker 2001), which could explain the non-spherical but
axisymmetric shapes that are so commonly observed (e.g., Balick
1987).  Rasio \etal\ (1996) have discussed the tidal decay of
planetary orbits during the red giant phase.  Indeed, even as far
back as 1924, Jeans examined the evolution of binary orbits due to
stellar mass-loss (only in that paper, the mass loss being considered
was in the form of the conversion of matter to energy via nuclear
fusion).  Moreover, several authors have considered the possibility
of detecting planetary companions to white dwarf stars, with the
goal of constraining the evolution of planetary systems through
empirical means (Zuckerman \& Becklin 1987; Livio, Pringle, \&
Saffer 1992; Li, Ferrario, \& Wickramasinghe 1998; Chu \etal\ 2001;
Ignace 2001; Burleigh, Clarke, \& Hodgkin 2002).  Recently, a
substellar companion to a subdwarf in a post-red giant phase of
evolution was reported (Geier \etal\ 2009).  This is particularly interesting
as an example of a low mass companion that avoided being engulfed
by the bloated red giant star to survive in a small period orbit
around the remnant.  Such occurrences bolster the need for the
consideration of detecting substellar companions to red giants to
understand better the connections between stellar evolution and
planetary evolution.

In this contribution one single fact is emphasized:  when stars
like the Sun
evolve to become red giants, the mass loss increases
substantially by several orders of magnitude which may have observational
consequences for detecting substellar companions and detailing their
physical and orbital properties.  Already the influence
of a strong wind on a substellar companion has been considered by
Struck, Cohanim, \& Willson (2003), who discuss the possibility of
wind accretion by brown dwarf companions during the asymptotic
giant branch (AGB) phase.

In terms of detecting extrasolar planets, a number of researchers
have explored radio emissions from substellar companions to stars
by extrapolating the radio properties of solar system planets to
other star systems (e.g., see Zarka 2007 and references therein).
The radio emission of the Earth and other solar system planets with
significant magnetic fields are dominated by the cyclotron maser
instability (CMI) process in which the low frequency spectrum
below about 100~MHz is dominated by coherent cyclotron emission
from mildly relativistic electrons (Gurnett 1974).  A key point is
that electrons are fed to the magnetic poles where the CMI is strong,
and that the coherent cyclotron emission dominates the incoherent
emission by several orders of magnitude.  A greater understanding
of the detailed energetics and emissive mechanism has come about
fairly recently in relation to the terrestrial auroral kilometric
radiation (AKR) by Mutel \etal\ (2007) and Mutel, Christopher, \&
Pickett (2008). 

In applications to extrasolar planets, Jupiter has been prominent in
these considerations, since it shows bursting behavior in the radio band
that can be quite bright, partly in relation to interactions occuring
between Jupiter and Io (e.g., Zarka 1998, 2004).  Griessmeier, Zarka,
\& Spreeuw (2007) have summarized various mechanisms that may contribute
to radio emissions in the case of exoplants around main sequence stars.
For example, coronal mass ejections and in some cases magnetic field
interactions between the star and a short-period planet can produce strong
radio signals.  Our contribution is to extend the use of the radiometric
``Bode's Law'' for late type main sequence stellar winds to those of giant
stars for which several new considerations must be taken into account.
This law is an empirically determined, and theoretically motivated,
power-law relation between planetary radio emissions and the solar
wind kinetic energy flux.  The radio emissions occur at low frequencies
of about 0.3--40~MHz, of which the upper end is observable with LOFAR
(Farrell \etal\ 2004). A lunar radio observatory such as the proposed
Dark Ages Lunar Interferometer (DALI) would be sensitive to almost that
entire range (see Lazio \etal\ 2007 for a summary of the capabilities
of lunar based radio observatories).

Giant star winds differ from those of low-mass main sequence
stars (i.e., solar-type winds) in important ways.  Giants that are
warmer than early K tend to be X-ray emitters, whereas those
that are cooler are X-ray faint.  The distinction in the X-ray
properties is known as the Linsky-Haisch ``dividing line'' (Linsky
\& Haisch 1979).  Although these X-ray faint giants possess
chromospheres, their winds are largely neutral (e.g., Reid \& Menten 1996),
whereas the earlier giants and the winds of low-mass main sequence
stars are ionised plasma flows.  Consequently, in the case of the
neutral giant star winds, the flow can {\em penetrate} the
magnetosphere of a substellar companion, and the supersonic wind
flow will set up a bow shock in the vicinity of that companion.
Cassinelli \etal\ (2008) have considered a closely related scenario.
Interested in X-ray emissions, those authors used hydrodynamic
simulations to model the bow shock for an early-type stellar wind
interacting with a dense spherical gaseous clump.  With a wind speed of order
$10^3$ km s$^{-1}$,
and assuming adiabatic cooling, they were able
to derive a power-law dependence for the differential emission
measure with temperature of the form $dEM/dT \propto T^{-7/3}$,
with peak hot gas temperatures of order $10^7$~K.

That simulation remains relevant for the case of a substellar companion
to a cool giant star.  Although cool star
winds are far slower than hot star winds, with values of 15 km~s$^{-1}$
for AGB stars up to perhaps 100 km~s$^{-1}$ for
some red giants, the winds are themselves much cooler, with sound speeds
of a few km~s$^{-1}$.  As a result, the wind is still quite supersonic.

However, this does not mean that bow shocks to substellar companions
will produce X-ray emissions for the cooler giant star systems.
These winds are indeed much slower than those of early-type stars,
and so the peak post-shock temperature will be considerably
smaller.  For a strong shock, as considered by Cassinelli \etal,
the post-shock temperature at the bow shock head (or ``apex'') $T_A$
follows the well-known Rankine-Hugoniot relation, with

\begin{equation}
T_A = \frac{3}{16}\,\frac{\mu m_H}{k}\,v_{\rm rel}^2 = 14\,{\rm MK}\,
	\left(\frac{v_{\rm rel}}{1000~{\rm km\; s^{-1}}} \right)^2.
	\label{eq:Ta}
\end{equation}

\noindent where $\mu$ is the mean molecular weight for the gas,
$m_H$ is the mass of hydrogen, and $v_{\rm rel}$ is the relative
speed between the wind and the companion.  The orbital motion can
be non-trivial compared to the wind, but considering only the wind
speed for the moment, one expects apex temperatures on the order
of $T_A \approx 3,000 - 140,000$ K for speeds of 15--100 km~s$^{-1}$.
At the low end, orbital motion will likely dominate the incident
shock speed in the rest frame of the companion (e.g., the Earth
orbits at about 30 km~s$^{-1}$).  As a result, post-shock temperatures
of $\sim 10^4-10^5$~K are expected, sufficient to ionise hydrogen,
thus leading to a substantial reservoir of electrons that can
interact with a magnetosphere to produce radio emissions.

The scenario for the giant stars is thus significantly different
than for main sequence stars.  First, known low-mass companions to
giants have no examples of extremely short period orbits (a week
or less) as in the main sequence case.  Second, giants with the
most massive winds will be neutral, and thus have no counterpart
in the solar system or to current applications among late-type main
sequence stars.  Third, for the X-ray emitters (or ``coronal''
giants), there has been no evidence for coronal mass ejections as
in the solar case, so that the application of the radiometric Bode's
law is likely the most important emissive mechanism in the long
wavelength radio band.  It is worth noting that Griessmeier \etal\
(2007) have evaluated the expected radio emissions for all known
exoplanets at that time.  Their list includes companions to giant
stars that had been reported at that time, and these do not produce
detectable radio emissions.  However, those authors appear to have
applied a main sequence solar wind model in relation mass-loss rate
and wind speed which is inapplicable to giant star winds.
Here we use empirical relations that are more appropriate for
evolved cool star winds.

To explore the radio emissions from companions to giant stars, the
next section summarizes the properties of the host stars and the
orbital properties of their companions.  In Section~3, we apply the
radiometric Bode's law to giant star systems, comparing emissions
between stars on either side of the Linsky-Haisch dividing line.
In Section~4, we consider the bow shock scenario for the case of neutral
winds that penetrate deep into a magnetosphere.  We use the results of
Cassinelli \etal\ (2008) to estimate the incoherent cyclotron emission.
Results of our study are summarized in Section~5, where we conclude that
new simulations are required to consider properly the operation of the
CMI in the bow shock scenario.  An Appendix details our derivation for
the cyclotron emission from a bow shock.

\begin{table*}
 \centering
 \begin{minipage}{140mm}
  \caption{Nominal Stellar and Planetary Data on Evolved Stars with Planetary Companions.}
  \begin{tabular}{@{}lcccccccccr@{}}
  \hline
        & \multicolumn{4}{c}{\it Host Star Data}  & &\multicolumn{4}{c}{\it Planet Data} \\
   Name & Spectral & Distance & Mass & Radius
     & Luminosity & $\dot M$ & $M\sin i$& $a$ & $e$ & Ref.$^a$ \\
        & Type  & (pc)  & $(M_\odot)$  & $(R_{\odot})$  & $(L_{\odot})$ & $(10^{-9} \,M_\odot$ yr$^{-1})$ & $(M_J)$  & (AU)  &       \\
        &  &  &  &  &  &       &     &       \\
 \hline
HD185269       &   G0 IV       &  47     &  1.28   &  1.88  &   4.0$^b$  &  0.002 &  0.94       &   0.077   &  0.30     &  J06     \\	 
 81 Cet        &   G5 III:     &   97    &   2.4   &   11   &   60    &  0.11 &   5.3    &   2.5    &   0.206   &   S08b \\	 
HD 11977       &   G5 III      &  67     &  1.91   &  10    &   55$^b$  &  0.12 &  6.54       &   1.93    &  0.4      &  Se05    \\	 
 18 Del        &   G6 III      &   73    &   2.3   &   8.5  &   40    &  0.059 &   10.3   &   2.6    &   0.08    &   S08a \\	 
 11 Com        &   G8 III      &   112   &   2.7   &   19   &   170   &  0.48 &   19.4   &   1.29   &   0.231   &   L08 \\	 
HD175541       &   G8 IV       &  128   &  1.65    &   3.80  &   8.6     &  0.008 &    0.61    &    1.03    &   0.33     &  J07     \\	 
HD192699       &   G8 IV       &  67     &   1.68  &   3.9   &   10.2     &  0.009 &    2.5    &    1.16    &   0.149    &  J07     \\	 
HD 104985      &   G9 III      &  102    &  1.6    &  11    &   59    &  0.16 &   63     &  0.78    &    0.03   &   Sa03    \\	 
$\xi$ Aql       &   K0 III      &   63    &   2.2   &   12   &   69    &  0.15 &   2.8    &   0.68   &   0$^d$      &   S08a \\	 
$\epsilon$ Tau  &  K0 III      &  47.5   &   2.7   &   14   &   97    &  0.20 &   7.6    &   1.93   &   0.151   &   S07 \\	 
HD102272       &   K0 III      &  360    &   1.9   &  10.1  &   53$^b$  &  0.11 &  5.9        &   0.61       &  0.05   &  N09b    \\	 
 14 And        &   K0 III      &   76    &   2.2   &   11   &   58    &  0.12 &   4.8    &   0.83   &   0$^d$      &   S08b \\	 
HD17092        &   K0 III      &  110    &   2.3   &  10.1  &   43$^b$     &  0.076 &  4.6        &   1.3       &  0.17    &  N07     \\	 
$\beta$ Gem    &   K0 III      &  10.3   &   1.86  &   9    &   34$^b$   &  0.066 &    2.9    &   1.69   &   0.06      &  H06, R06     \\	 
HD81688        &   K0 III-IV   &   88    &   2.1   &   13   &   72    &  0.18 &   2.7    &   0.81   &   0$^d$      &   S08a \\	 
$\kappa$ Cr B  &   K0 IV       &  31.1   &   1.8   &   4.71  &   12.3      &  0.013 &   1.8    &    2.7     &   0.146    &  J08    \\	 
 6 Lyn         &   K0 IV       &   57    &   1.7   &  5.2   &   15    &  0.018 &   2.4    &   2.2    &   0.134   &   S08b \\	 
HD32518        &   K1 III      &   120   &   1.13  &  10.2  &   41$^b$  &  0.15 &  3.04  &  0.59     &   0.01      &   D09b\\	 
4 U Ma         &   K1 III      &  62     &  1.23   &  18    &   110$^b$   &  0.64 &  7.1        &   0.87    &  0.43     &  D07     \\	 
HD167042       &   K1 III      &  50.    &   1.64  &   4.30  &   10.5     &  0.011 &    1.7    &    1.3     &   0        &  J08     \\	 
HD 47536       &   K1 III      &  121    &   1-1.5  &  23   &   4380   &  40 &   5-6   &   2     &    0.20      &   Se03       \\	 
HD167042       &   K1 IV       &   50    &   1.5   &   4.5  &   10    &  0.012 &   1.6    &   1.3    &   0.101   &   S08b \\	 
$\gamma$ Ceph  &   K1 IV       &  13.8   &   1.6   &   4.66  &   11$^b$  &  0.013 &   1.7     &   2.13   &   0.12     &  H03     \\	 
HD210202       &   K1 IV       &  56    &  1.85    &   4.45  &   11.3     &  0.011 &    2.0    &    1.17    &   0.152    &  J07     \\	 
42 Dra         &   K1.5 III    &   97    &   0.98  &   22   &   130$^b$  &  1.2 &  3.88  &  1.19    &  0.38       &  D09a \\	 
BD +20 2457    &   K2 II       &   200+$^c$  &  2.8  &  49  &   610$^b$     &  4.3 &  21.42  &  1.45  &  0.15   &  N09a \\	 
BD +20 2457    &   K2 II       &   200+$^c$  &  2.8  &  49  &   610$^b$     &  4.3 &  12.47  &  2.01  &  0.18   &  N09a \\	 
HD 13189       &   K2 II?      &  --- &  2-7    &   --- &   4000   &  --- &   8-20   &  1.5-2.2   &   0.27    &   H05     \\	 
$\iota$ Dra    &   K2 III      &  31     &  1.05   &  12.9  &   70    &  0.34 &   8.9    &   1.3    &    0.70   &   F02    \\	 
HD24210        &   K3 III      &   140   &   1.25  &  56    &   950$^b$  &  17 &  6.90  &  1.33  &  0.15    &  N09 \\	 
HD139357       &   K4 III      &   118   &   1.35  &  11.5  &   58$^b$   &  0.20 &  9.76  &  2.36    &  0.10       &  D09a \\	 
 11 U Mi       &   K4 III      &   120   &   1.8   &   24   &   180$^b$  &  0.96 &  11.2  &  1.54     &   0.08    &   D09b\\	 
\hline
\end{tabular}
$^a$References:  D09 = Dollinger et al. (2009b);  
D09a = Dollinger et al. (2009a);
N09a = Niedzielski et al. (2009a);
L08 = Liu et al. (2008);
S08a = Sato et al. (2008a);
S08b = Sato et al (2008b);
S07 = Sato et al (2007)
N09b = Niedzielski et al. (2009b);
N07 = Niedzielski et al. (2007);
D07 = Dollinger et al. (2007);
J06 = Johnson, et al. (2006);
Se05 = Setiawan et al. (2005);
H05 = Hatzes et al. (2005);
Sa03 = Sato et al. (2003);
F02 = Frink et al. (2002);
Se03 = Setiawan et al. (2003);
H06 = Hatzes et al. (2006);
R06 = Reffert et al. (2006);
H03 = Hatzes et al. (2003);
J07 = Johnson et al. (2007);
J08 = Johnson et al. (2008). \\
$^b$ Luminosity computed from values of $T_{\rm eff}$ and $\log g$ given by respective reference. \\
$^c$ Luminosity estimate based on spectral type. \\
$^d$ Eccentricity fixed to zero for orbital solution. \\

\end{minipage}
\end{table*}
\section{Giant Stars and Their Substellar Companions}

A large number of substellar companions have been identified in
radial velocity surveys of giant stars.  From the literature we
have gathered information about their physical properties
which are displayed in Table~1. 
Distances were taken from the SIMBAD database that lists values
from the Hipparcos survey.  
References for the stellar and planetary properties appear in the last
column.  Note that some of the values in the table are quite uncertain, and
the reader is strongly urged to consult the references for further
details of their evaluation.  Our goal here is to use these values as
a guide for our application to the radio emissions.

With these stellar parameters we use a
standard form of Reimer's law to estimate wind mass-loss rates
$\dot{M}$ (Reimers 1975; Lamers \& Cassinelli 1999):

\begin{equation}
\dot{M} = 4\times 10^{-13} \,\eta\,\frac{(L_\ast/L_\odot)\,(R_\ast/
	R_\odot)}{(M_\ast/M_\odot)}\, M_\odot\;{\rm yr}^{-1},
\end{equation}

\noindent where $\eta$ is a scaling parameter between about 0.3 and 3.
Here, we adopt $\eta=1$ since only mass-loss rate estimates will be
needed.  Computed mass-loss rate values are given in Table~1 in the
seventh column, scaled to $10^{-9} \, M_\odot$ yr$^{-1}$.  Note that
no value is given for HD13189 since no radius or temperature value
was quoted for this star.

The mass loss is important since it sets the scale of the wind
density at the companion where radio emission will be produced.
Table~1 shows that there is a significant spread in values fro
the giants with planets detected so far, ranging from
$4\times 10^{-8}\, M_\odot$ yr$^{-1}$ for HD47536 down to about
$10^{-12}\, M_\odot$ yr$^{-1}$ for the subgiant HD185269.
However, there is controversary about Reimer's law and its applicability.
An understanding of it remains a topic of current research (e.g.,
Schr\"{o}der \& Cuntz 2005).  The law has application to more luminous
cool stars, but its extension to their lower luminosity cousins is
less clear.  Also ongoing are attempts to understand the Linsky-Haisch
``dividing line'' that separates giant stars earlier than about K0
that show X-ray emissions from the later types that are not coronal 
but which do have chromospheres (Suzuki 2007).
Again, the main value of Table~1 is the determination of representative
stellar and planetary parameters and their spread for the sample as a whole.

In relation to the population of host giant stars, the notable points are that
the median distance is around 100~pc, about 10 times farther than main
sequence host stars of substellar companions.  Also, the mass-loss rates
are typically of order $10^{-10}\ M_\odot$ yr$^{-1}$, about 4 orders
of magnitude larger than the solar wind, but lower by a similar factor
from the AGB winds.  Finally, the distribution of orbital semi-major
axes and eccentricities reflects the selection effects of the
radial velocity study:  values of $a$ around 1~AU or less, and
values of $e$ that can deviate significantly from circular orbits.

\section{Application of the Radiometric Bode's Law to Giant Star Winds}

Radio studies of single red giant stars reveal them to be faint radio
emitters (Spergel, Giuliana, \& Knapp 1983; Reid \& Menten 1996;
G\"{u}del 2002).  For the brighter sources, observations at different
radio bands indicate a flux density spectrum of the form $S_\nu
\propto \nu^2$, consistent with the Rayleigh-Jeans limit and suggestive
of photospheric emission, although there are some exceptions
indicating extended radio photospheres.

The main conclusion is that the red giant star winds that show
chromospheric signatures but not coronal X-rays are largely neutral,
yet some metals are ionised, resulting in low level ionisations of
the wind material at the level of
0.01-0.1\% (Reid \& Menten 1996).  These
ionisations are 3 to 4 orders of magnitude below that of the Sun's
wind.  However, the mass-loss rate from giant stars is 4 or more
orders of magnitude larger than the solar wind.  Interestingly,
Judge \& Stencel (1991) have reported on larger ionisations
in some red giant stars.  It is true that the earlier red giants,
earlier than about K2 that show X-ray emissions, have fully ionised
winds.  Unfortunately, their mass-loss rates are relatively uncertain.
It is useful to consider how even the paltry ionised
component of a giant star wind might interact with planetary
magnetospheres so that at least lower limits to the radio emissions
may be derived.

The topic of radio emissions in the solar system and from extrasolar
planets has been studied extensively.  To estimate the flux density
from the giant star wind impinging on a planetary magnetosphere, we
use the radiometric ``Bode's'' law of
equation (3) from Lazio \etal\ (2004) based on scaling
relations from Farrell, Desch, \& Zarka (1999) and Zarka \etal\ (2001).
The median radio power from the stellar wind interaction with the
planet is 

\begin{eqnarray}
L & \approx & 4\times 10^{18}\;{\rm erg\;s^{-1}}\,\left(\frac{\omega}{\omega_J}
	\right)^{0.79}\,\left(\frac{M}{M_J}\right)^{1.33}\,\left(
	\frac{d}{5~AU}\right)^{-1.6} \, \nonumber \\
 & & \times \left(\frac{\rho\,v_{\rm w}^3}
	{\rho_\odot\,v_\odot^3}\right),
	\label{eq:Lrad}
\end{eqnarray}

\noindent where $\omega$ is the planet's rotation rate, $M$ is its
mass, $d$ is the orbital distance of the planet from its star, $\rho$
and $v_{\rm w}$ are the density and speed of the stellar wind, 
$\rho_\odot$ and $v_\odot$ are the density and speed of the solar wind,
and ``J'' subscripts indicate values for Jupiter.
The density $\rho$ must be corrected for the lower ionisation level
of a late giant star wind as compared to a main sequence star.
Introducing $\eta$ as a ratio of stellar wind
properties to that of the Sun, and assuming a spherically symmetric
wind with $\rho = q\dot{M}/4\pi\ d^2v_{\rm w}$, we can derive that

\begin{equation}
\eta = \frac{q\dot{M}\,v_{\rm w}^2}{\dot{M}_\odot\,v_\odot^2},
\end{equation}

\noindent where $q$ is the ionisation correction factor, with a value of
about $10^{-4}$ to $10^{-3}$.  With $\dot{M}_\odot \approx 10^{-14}\,
M_\odot$ yr$^{-1}$ and $v_\odot \approx 400$ km~s$^{-1}$ for the Sun,
and $\dot{M} \approx 10^{-8}\, M_\odot$ yr$^{-1}$ and $v_{\rm w} \approx
30$ km s$^{-1}$ for a giant star wind with $q = 10^{-3}$, we arrive
at $\eta \sim 6$ for a red giant with fairly strong mass loss.  Some
red giants have lower $\dot{M}$ values and faster speeds, in which
case $\eta$ would remain comparable to the above value.  AGB stars
on the other hand are slower by a factor of two but have higher
mass loss by a factor of $10^3$, and so $\eta$ can become quite
large at around $10^3$.

Assuming $\eta \sim 10$, and the other scaling ratios are unity
in equation~(\ref{eq:Lrad}), then the total radio luminosity would
be approximately $4\times 10^{19}$ erg s$^{-1}$.  Using equation~(4)
of Lazio \etal\ for a typical emission frequency of $\nu \sim 25$
MHz, combined with their equation~(5) for the flux density under
the approximation that the radio luminosity is emitted isotropically
in a bandwidth of $\Delta \nu \approx \nu/2$, the expected radio
brightness level will be

\begin{eqnarray}
S_\nu & \approx & \frac{L}{4\pi\,(\nu/2)\,D^2} \nonumber \\
 & \approx & 1.0 \,\mu{\rm Jy}\,\frac{q}{10^{-3}}\,
	\left(\frac{\omega}{\omega_J}
	\right)^{-0.21}\,\left(\frac{M}{M_J}\right)^{-0.33}\,\left(
	\frac{d}{1~AU}\right)^{-1.6}
	\nonumber \\
 & & \times \left(\frac{\nu}{12.5~{\rm MHz}}\right)^{-1}\,
	\left(\frac{\Omega}{4\pi}\right)^{-2}\,
	\left(\frac{D}{100~{\rm pc}}\right)^{-2}\,\left(\frac{\dot{M}}
	{10^{-8}}\right)\,\nonumber \\
  & & \left(\frac{v_{\rm w}}{30}\right)^2\,\left(
	\frac{R}{R_J}\right)^{-3},
	\label{eq:snu}
\end{eqnarray}

\noindent where $D$ is the Earth-star distance, 
and $\Omega$ is the beaming of the radiation
relative to isotropic.  If all of the parenthetical factors were unity,
then a Jupiter-like companion to a red giant star
at 100~pc would have a flux density of
$S_\nu \approx 1~\mu$Jy.  This is too faint for detection by current
or near-future radio telescopes.  

However, it should be noted that
we have assumed nearly neutral winds, referring to late red giants,
those that lie on the cool side of the Linsky-Haisch dividing line.
Red giants of earlier spectral type
that are X-ray emitters likely have winds similar
to that of the Sun, namely fully ionised plasmas.  For such stars
$q \approx 1$, with a gain factor in the radio emission of 3 orders
of magnitude, bringing the flux density up to about 1~mJy, which is
within the realm of detection by facilities such as LOFAR.

Unfortunately, most giant stars that are known host stars for substellar
companions have much lower mass-loss rate values than assumed in
equation~(\ref{eq:snu}), closer to $10^{-10} M_\odot$ yr$^{-1}$.
For an ionised red giant wind with $q=1$, this level of mass loss
pushes the expected flux density down to about 10~$\mu$Jy, again well
below detection thresholds.  However, some G~giants can have $\dot{M}$
values closer to $10^{-9} M_\odot$ yr$^{-1}$ (e.g., 11~Com in Tab.~1).
It is easy to imagine favorable cases with an early giant star wind with
somewhat large mass loss and higher wind speed along with a companion
in a sub-AU orbit (possibly eccentric -- see Sect.~5) that could return
the radio flux density closer to the mJy level.

In addition, it is possible that in fact the majority of giant stars
maintain planetary systems or brown dwarf companions.  Substantially
higher radio flux densities would then be expected from giants with
higher mass-loss rates, ones that have not been included in the radial
velocity surveys.  For example, AGB stars with $\dot{M} \gtrsim 10^{-7}
M_\odot$ yr$^{-1}$ could represent a new stellar population for detecting
substellar companions in later stages of stellar evolution than has so
far been targeted. \\

\section{Radio Emission from a Wind Bow Shock}

We now turn to a new consideration for generating radio emissions.
Since some giant star winds are neutral, they will penetrate
companion magnetospheres and can be expected to set up bow shocks in
close vicinity of the substellar object.  Cassinelli \etal\ (2008)
describe a hydrodynamic simulation of a massive star wind impinging on
a spherically symmetric ``hard'' clump.  For these authors the focus
was on explaining X-ray emissions from hot star winds.  But the key
results of the simulations pertain to emission measure and temperature
distributions, and so the results have relevance for bow shocks formed
from red giants that intercept substellar companions.  To explore the
observational consequences of this scenario, we adopt expressions for the
emission measure ($EM$) distribution of post-shocked gas as a function
of temperature ($T$) from equations (23) and (24) of Cassinelli \etal\
(2008):

\begin{equation}
\frac{dEM}{dT} = \frac{EM_0}{T_A}\,\left(\frac{T}{T_A}\right)^{-7/3},
	\label{eq:demdt}
\end{equation}

\noindent where $T_A$ is given from equation~(\ref{eq:Ta}).  The
emission measure scaling factor $EM_0$ is set by the square of the wind
number density (times four owing to a strong shock) at the location of
the companion multiplied by the volume of the companion.  Its value is
given by

\begin{equation}
EM_0 = 1.4\times 10^{46}\,{\rm cm^{-3}}\,\left(\frac{R}{10^{10}}\right)^3\,
	\left(\frac{\dot{M}_{-8}}{\mu\,v_{30}\,r_{AU}^2}\right)^2,
\end{equation}

\noindent with $R$ the planet radius, $\dot{M}_{-8}$ the wind
mass-loss rate divided by $10^{-8}\, M_\odot$ yr$^{-1}$, $v_{30}$
the radial wind speed divided by 30 km~s$^{-1}$, and $r_{AU}$ the
orbital distance of the planet in AU.  Notice that the EM is a
strong function of the planet size (as the cube) and the orbital
distance (as the fourth power).  The sizes of Jovian planets and
brown dwarfs vary slowly with radius, except for short period
companions where X-ray and UV heating and tidal effects can enlarge
the effective radius of the planet (e.g., Guillot \etal\ 1996;
Lammer \etal\ 2003; Burrows \etal\ 2007).  We assume a
nominal value of $R\approx 10^{10}$~cm.

The scenario that we envision is one where the largely neutral wind
penetrates a planetary or brown dwarf magnetosphere.  On the scale
of the companion size , the wind flow is approximately plane parallel
as in the Cassinelli \etal\ (2008) simulation for a wind clump.
The peak temperature achieved at the bow head will be of order
$10^4$~K or more.  The hydrogen gas can become ionised, and the bow
shock has a decreasing temperature distribution along its length
downstream.  The shock becomes increasingly oblique until it drops
down to around 3,000~K, where we assume that hydrogen is no longer
ionised at the shock front.

This reservoir of electrons finds itself embedded deep inside
a magnetosphere that is sweeping past them.  Ignoring the details of
the flow dynamics, we make a lower limit estimate for
the radio emission as arising from
a non-relativistic, incoherent cyclotron process.  This emission will occur
at a characteristic frequency of

\begin{equation}
\nu_{\rm c} = \frac{e\,B}{2\pi\,m_{\rm e}\,c},
\end{equation}

\noindent which is about 28~MHz for a field strength of 10~G.
In fact, the emission occurs within the volume of the bow shock,
predominantly in a higher density region that hugs the bow shock
itself (see Cassinelli \etal\ 2008).  A derivation for the
flux density and spectral shape of the thermal, non-relativistic
cyclotron emission is given in the Appendix.  The result is 
reproduced here (see eq.~[\ref{eq:SnuBow}]), with the flux density

\begin{eqnarray}
S_\nu & \sim & 0.3\ \mu Jy\,\left(\frac{B_0}{30~G}\right)\,
        \left(\frac{T_A}{30,000~K}\right)\,
        \left(\frac{EM_0}{1.4\times 10^{46}}\right)\,\nonumber \\
  & &   \left(\frac{n_{\rm e}}{5.7\times 10^7}\right)^{-1} 
	\left(\frac{D}{100~pc}\right)^{-2}\,
        \left(\frac{\nu}{\nu_A}\right)^{1/3}\,
        x^{2/3}\,\xi^{-1},
\end{eqnarray}

\noindent where $x=T/T_A$, and $\xi=\xi(x)$ as described in
equations~(\ref{eq:xi}) and (\ref{eq:y}) of the Appendix.  The scale
of this emission is sub-microJansky, well below the detectability
of current instrumentation.  However, there are several key points
to note.  First and foremost, the above calculation should be
considered as a minimum flux density in the sense that it does not
take account of CMI, or any bursting behavior such as is observed
in Jupiter, and such emissive processes will be stronger by orders of
magnitude over the thermal cyclotron emission that we have considered.
Nor does it account for the possible influence of moons, each of
which would have its own bow shock.  (Of course, being a smaller
target, the overall EM from a moon's bow shock would
be much smaller than for a Jovian-like object, yet the shock
could have hotter gas owing to its circumplanetary orbital motion.)
The ionisation is set only by the relative flow between the blunt
object and the wind.  
Although the EM of the bow shock for
a moon would be insignificant as compared to the companion bow
shock, its primary influence may be in the form of providing an
injection mechanism of electrons to the polar field regions of the
companion where the CMI operates.

It should be noted that much of what is being proposed for the bow
shock is qualitative only.  For a magnetosphere with a typical field
strength of order 10~G in the vicinity of the bow shock, the magnetic
energy density $U_B = B^2/8\pi \sim 4$ erg cm$^{-3}$ is vastly larger
than the ram pressure of the wind flow $\rho v^2 \approx 10^{-3}$
erg cm$^{-3}$.  The implication is that a very strong field is rotating
past the bow shock, that in essence acts as an ionisation front.  The
post-shocked plasma is essentially being ``slammed'' by the magnetosphere
and rapidly accelerated.  It is unclear what effect this will have on the
bow shock structure itself.  A fully consistent MHD simulation of this
scenario needs to be carried out .  It is a situation unlike anything
occuring in the solar system where the solar wind is everywhere extremely
fast and highly ionised in its interactions with solar system bodies.
Importantly, it is unclear in the neutral giant wind case how or if
electrons can be accelerated to mildly relativistic values and also
channeled to the polar regions of the field axis for the CMI to operate.

However, it is encouraging that there is current interest in both
theoretical and experimental work for understanding the CMI process.
For example, McConville \etal\ (2008) conducted a laboratory experiment
for a scaled version of the terrestrial AKR that yielded results in
basic agreement with satellite measurements.  Their work revealed that
approximately 1\% of the electron kinetic energy pool was converted to
radiation via the CMI.  These results appear reproducible in 3D numerical
simulations (Gillespie \etal\ 2008).  Such work supports the ansatz that
a similar level of efficiency can be used in applications to extrasolar
planets (e.g., Jardine \& Cameron 2008).  To advance modeling of the
radio emissions in the red giant wind case, MHD simulations will be
necessary to assess how electrons can be fed to the magnetic poles of
the companion, and then combine that information with the body of work
just described for the AKR mechanism.  It seems likely that such an
approach will be tractable.

\section{Discussion \label{sec:disc}}

On the whole our analysis for the detection of radio emissions
from the interaction between the wind of a red giant star
and its substellar companion is largely negative.  Cyclotron
emissions appear to be at the micro-Jansky level, orders of
magnitude below current or near-future detection thresholds.
However, we have taken the most conservative and pessimistic
approach in our estimates.  We have considered red giant
winds that are largely neutral (i.e., those on the cool side
of the ``dividing line''), yet the winds of earlier type red
giants may be highly ionised with large gain factors in
the resultant radio emissions.  

There have been radio-wavelength surveys (typically at 2 and 6 cm)
of red giant stars including stars on the asymptotic
giant branch of the H-R diagram (see Drake \& Linsky 1983, 1986;
Drake, Linsky, \& Elitzur 1987; Drake \etal\ 1991; Luttermoser \&
Brown 1992).  Typically evolved stars on the cool side of the
``dividing line'' are either not detected at these wavelengths or
are very weak sources --- giant stars of M6 or warmer have been
detected (Drake \etal\ 1991).  The weak radio emission of these
non-Mira giants are thought to be due to partially ionised winds.
There have been some detections of giants on the cool side of M6~III.
For example $o$~Ceti (Mira) appears variable in its radio emission.
Mira was detected at 6~cm by Spergel, Giuliana, \& Knapp (1983) as
a $3\sigma$ event; however, Drake \etal\ (1987) failed to
detect this pulsating long-period variable (LPV) at either 2~cm or
6~cm.  Since Mira has a hot companion, it
is not clear if the emission arose from the LPV or from the companion.
In a survey of seven N-type carbon stars at 3.6~cm, Luttermoser \&
Brown (1992) only detected V~Hya, which is a peculiar carbon star
that shows evidence of bipolar outflow (Tsuji \etal\ 1988).

In the case of a neutral wind, we have argued that a bow shock
will result in the vicinity of the companion object.  If the
relative flow speed is enough, one can expect post-shock
ionisation of the wind material.  For incoherent cyclotron emission, the
radio flux of the ionised component will be quite weak.  However,
it may be possible that some fraction of these newly created
electrons are fed to the polar regions of the companion where
the CMI mechanism can operate leading to much stronger radio
emission.  In addition, analogous to the Jupiter-Io interaction,
moons of substellar companion may also produce bow shocks,
and these might provide a source of electrons that could be accelerated
in the magnetosphere and feed the CMI.
Our model for the bow shock is not adequate to address these
possibilities.  New self-consistent MHD calculations will
be needed to model the interaction of the bow shock with
the companion's magnetosphere and to assess the channeling of
electrons for use with the CMI.

Of course, all of the effects described so far would be affected by
orbital eccentricity of the companion.  For orbits of a fixed semi-major
axis, the radio flux density will be significantly higher near periapse as
compared to apapse for increasingly eccentric orbits.  But then of course,
gains in signal strength near periapse will occur over a diminishing
fraction of the orbital period as eccentricity increases.  Although
these effects should be explored, they are of secondary importance
to substantial task of modeling the flow dynamics more accurately.

Finally, it is possible that the AGB stars might be targeted for long
wavelength radio studies.  The AGB winds are certainly neutral so that
the bow shock scenario would be of interest.  The mass-loss rates can
be extremely large at around $10^{-5}\,M_\odot$ yr$^{-1}$.  Since the
radio emission scales with $\dot{M}^2$, there would be a gain factor
of 6 orders of magnitude above our estimates for the radio emission,
and that is just for incoherent cyclotron emission.  Of course, AGB
stars are relatively rare and so tend to be more distant.  Plus the
companion will probably need to be in an orbit that is greater than
1~AU, otherwise it could be engulfed by the star.  With a wind speed
of just 15 km s$^{-1}$, the astute reader may recognize that the apex
temperature will hardly be large enough to ionise hydrogen.  However,
there are actually two effects to mitigate the low wind speed.  First is
that the orbital speed could be larger by perhaps a factor of 2 or more.
Second, the hydrodynamical simulations of Cassinelli \etal\ (2008) are
for clumps treated as blunt obstacles without self-gravity.  Jovian and
brown dwarf mass objects have surface escape speeds of 30 km s$^{-1}$
and higher.  Given that the bow shock apex is at around 1.5 times
the radius of the blunt object in the Cassinelli \etal\ simulations,
a slow AGB wind can double and triple in speed as it falls into the
gravitational potential of the substellar companion (i.e., similar to
accretion described by Bondi \& Hoyle 1944).  Of course, inclusion of
gravity changes the details of the bow shock, but the main point is that
significant ionisation would still be viable.  Collecting the various
factors, $S_\nu$ of around 0.1~mJy might be achievable.

It is clear that long wavelength observations will provide new
opportunities for detecting substellar companions and for studying their
magnetospheres.  Companions to red giants should not be neglected in these
efforts.  An interesting result of our study has been the relatively novel
consideration of neutral winds interactions with substellar companions,
completely unlike the solar system case.  The prospects for future
observations are certainly exciting, but there is a clear need for more
detailed modeling of MHD bow shocks in relation to the CMI to explore
the viability of radio detection.

\section*{Acknowledgements}

The authors are gratefully indebted to Bob Mutel for discussions that
clarified the AKR process as well as an anonymous referee whose comments
led to improvements of this paper.  This research has made use of the
SIMBAD database, operated at CDS, Strasbourg, France.

\appendix

\begin{figure}[h]
\resizebox{\hsize}{!}{\includegraphics[width=17cm]{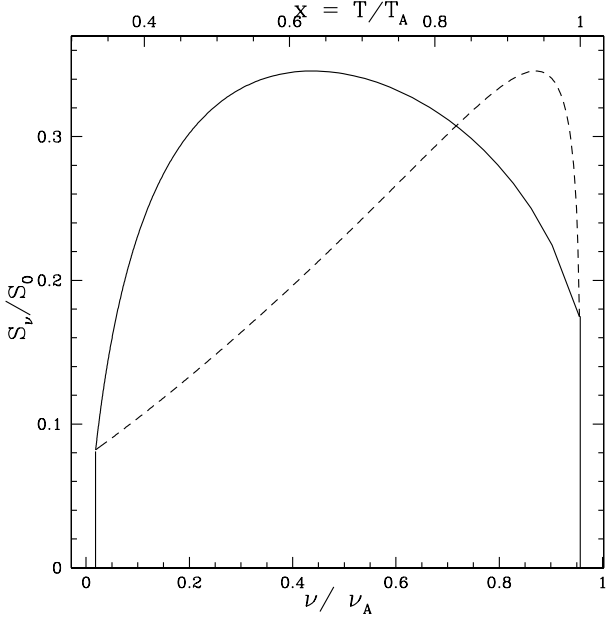}}
\caption{A plot of the radio spectrum with frequency (lower axis
and solid curve) and temperature (upper axis and dashed curve).
The temperature ranges from the apex value $T_A$ down to a lower
cut-off corresponding to the shock being too weak to ionise H, here
taken as $0.4T_A$.  Since the bow shock penetrates deepest into the
magnetosphere, the highest temperature point is likely to sample
the highest magnetic field value corresponding to emission at
frequency $\nu_A$.  The spectrum is illustrative only, as it assumes
$B \propto r^{-3}$ without taking any account of latitutindal
dependence.  \label{fig2}} 
\end{figure}

\section{Thermal Cyclotron Emission from the Bow Shock}

The steps for determining the radio flux density in the bow shock
model is detailed.
The differential luminosity $dL$ of cyclotron emission from a 
small unit of volume $dV$ is given by

\begin{equation}
dL = \frac{\sigma_T}{6\pi\, c}\,v_{\rm e}^2\,B^2\,n_{\rm e}\,dV,
\end{equation}

\noindent for thermal cyclotron emission by non-relativistic electrons.
For an environment with a position dependent magnetic field, the
frequency bandwidth of the emission is given by

\begin{equation}
d\nu = \frac{e}{2\pi\,m_{\rm e}\,c}\,dB.
\end{equation}

\noindent Thus the specific luminosity becomes

\begin{equation}
L_\nu = \frac{dL}{d\nu} = \frac{1}{6\pi}\,\frac{\sigma_T\,v_{\rm e}^2}
	{c}\,\frac{B^3}{\nu_{\rm c}}\,n_{\rm e}\, \frac{dV}{dB},
\end{equation}

\noindent where $\nu_{\rm c}$ is the cyclotron frequency.  For the
bow shock model of Cassinelli \etal\ (2008), the EM
$EM$ is dominated in large part by the region along the bow shock.
We can make the following approximation for the cyclotron emitting volume,

\begin{equation}
n_{\rm e}\,dV = \frac{1}{n_{\rm e}}\,dEM.
\end{equation}

\noindent This last expression takes advantage of the ``on the shock
approximation'' (OTSh) of Cassinelli \etal\ (2008), for which
$n_{\rm e} \approx constant$ along the bow shock in the strong shock
limit.  This occurs under the assumption that the extent of the bow
shock is relatively small compared to the orbital radius.

The specific luminosity of cyclotron emission now becomes

\begin{equation}
L_\nu = \frac{\sigma_T\,v_{\rm e}^2\,B^3}{6\pi\,n_{\rm e}\,c\,\nu_{\rm c}}\,
	\frac{dEM/dT}{dB/dT},
\end{equation}

\noindent where we have taken $dEM/dB$ as a ratio of parametric form
in the temperature $T$.  Since $T$ is a smoothly decreasing function
along the bow shock surface, it acts effectively as a mapping coordinate
in our prescription for the thermal cyclotron emission.  The differential
EM is simply a power law in temperature given by

\begin{equation}
\frac{dEM}{dT} = \frac{EM_0}{T_A}\,\left(\frac{T}{T_A}\right)^{-7/3}.
\end{equation}

In order to determine $dB/dT$, we require two pieces of information.
The first is how the field varies in space around the companion
object.  We will assume that the field is a dipole.  Since we
are seeking mainly an order of magnitude estimate of the emission
level at long wavelengths in the radio, we will ignore latitudinal
variations of the field and simply adopt $B = B_0 \, (R/r)^3$, where
$R$ is the companion radius.  The second piece is the mapping
of $r(T)$ along the bow shock as it cuts through the dipole field.
From Cassinelli \etal\ (2008), the bow shock shape is close
to a parabola (especially near the bow head, that is
most relevant for our work), which is a convenient form to work with.  Again,
with the goal of obtaining an order of magnitude estimate, we
approximate the bow shock geometry using

\begin{equation}
z = z_0 + a\,R\,\left(\frac{\varpi}{R}\right)^2,
\end{equation}

\noindent where $z_0 =1.19 R$ and $a= 0.35$, where $(\varpi,z)$ are
cylindrical coordinates for the axisymmetric bow shock with $z$ the
symmetry axis.  The way to obtain $B(T)$ is a two step process.
The first step is to derive $r(\varpi)$; the second is to use
$\varpi(T)$ from Cassinelli \etal.

With $r^2 = z^2 + \varpi^2$, the solution for $r(\varpi)$ is derived
from a quartic relation in $\varpi(r)$ that can be inverted to obtain
$r$ itself; the result is

\begin{equation}
\frac{r(\varpi)}{R} \approx \sqrt{1.4 + 5.7\varpi^2+1.1\varpi^4}.
\end{equation}

\noindent The solution for $\varpi(T)$ is approximately given by

\begin{equation}
\frac{\varpi(T)}{R} \approx \left[\frac{3}{2}\,\left(\frac{T_A}{T}-1\right)
	\right]^{3/8}.
\end{equation}

\noindent Finally, the end result for $dB/dT$ can be compactly expressed
as

\begin{equation}
\frac{dB}{dT} \approx 9 \,\left(\frac{R^2}{r^2}\right)\,\left(\frac{B}{T_A}
	\right)\, \xi\,x^{-2},
\end{equation}

\noindent where

\begin{equation}
x = T/T_A
	\label{eq:x}
\end{equation}

\noindent and 

\begin{equation}
\xi = \frac{1}{y^{1/4}} + \frac{1}{2}\,y^{1/2},
	\label{eq:xi}
\end{equation}

\noindent with

\begin{equation}
y = \frac{1}{x} - 1.
	\label{eq:y}
\end{equation}

For the electron velocity, we use the rms thermal value of

\begin{equation}
v_{\rm e} = \frac{2kT}{m_{\rm e}}.
\end{equation}

\noindent Combining the preceding relations, noting that the flux
density is $S_\nu = L_\nu /4\pi D^2$ for distance $D$, and
introducing a maximum frequency of emission $\nu_A$ corresponding
to the apex of the bow shock that temperature $T_A$ and minimum
radius $r_A = r(T_A)$, the flux density becomes

\begin{eqnarray}
S_\nu & \sim & 0.3\ \mu Jy\,\left(\frac{B_0}{30~G}\right)\,
	\left(\frac{T_A}{30,000~{\rm K}}\right)\,
	\left(\frac{EM_0}{1.4\times 10^{46}}\right)\,\nonumber \\
 & & 	\left(\frac{n_{\rm e}}{5.7\times 10^7}\right)^{-1} 
	\left(\frac{D}{100~{\rm pc}}\right)^{-2}
	\left(\frac{\nu}{\nu_A}\right)^{1/3}
	x^{2/3}\xi^{-1}.
	\label{eq:SnuBow}
\end{eqnarray}

\noindent The total emission sub-microJansky and peaks close to
the maximum frequency.  An example spectrum is shown in Figure~\ref{fig2}.
The flux density is normalized to 0.3~$\mu$Jy with nominal
values assumed for the physical parameters in equation~(\ref{eq:SnuBow}).
The lower temperature bound is taken as $T_0=10,000$~K.  The solid
curve is the for the frequency spectrum; the dashed curve plots the
emission against the temperature distribution with scale given at top.
It is assumed that hydrogen is completely ionised over this temperature
range.

\end{document}